\documentclass[12pt]{article}
\usepackage{epsfig}
\makeatletter
\def\alt{\mathrel{\mathpalette\gl@align<}}
\def\agt{\mathrel{\mathpalette\gl@align>}}
\def\gl@align#1#2{\lower.6ex\vbox{\baselineskip\z@skip\lineskip\z@
\ialign{$\m@th#1\hfil##\hfil$\crcr#2\crcr\sim\crcr}}}
\makeatother

\begin{document}
\begin{flushright}
{\tt hep-ph/0702002}\\
MIFP-07-04 \\
January, 2007
\end{flushright}
\vspace*{2cm}
\begin{center}
{\baselineskip 25pt \Large\bf
Landscape of Little Hierarchy
}

\vspace{1cm}

{\large
Bhaskar Dutta and Yukihiro Mimura}
\vspace{.5cm}

{\it
Department of Physics, Texas A\&M University,
College Station, TX 77843-4242, USA
}
\vspace{.5cm}

\vspace{1.5cm}
{\bf Abstract}
\end{center}

We investigate the little hierarchy between $Z$ boson mass and the
SUSY breaking scale in the context of landscape of electroweak
symmetry breaking vacua.
We consider the radiative
symmetry breaking and found that the scale where the electroweak
symmetry breaking conditions are satisfied
 and the average stop mass scale is
preferred to be very close to each other in spite of the fact
that their origins depend on different
parameters of the model.
If the electroweak symmetry breaking scale is fixed at about 1 TeV
by the supersymmetry model parameters then  the little hierarchy seems to be
preferred among the electroweak symmetry breaking vacua.
We  characterize the little hierarchy
 by a probability function and the mSUGRA model is used as an
example to show the 90\% and 95\% probability contours in the
experimentally allowed region.
We also investigate the size of the Higgsino mass $\mu$ by considering
the distribution of electroweak symmetry breaking scale.


\thispagestyle{empty}

\bigskip
\newpage

\addtocounter{page}{-1}

\section{Introduction}
\baselineskip 18pt

One of the key motivations for supersymmetric (SUSY)  extension of
the Standard Model (SM) is the stabilization of the large hierarchy
between the Planck scale and the weak scale.
Although the SM particle spectrum gets doubled in the SUSY extension,
these new particles around the weak scale add an additional attraction for
SUSY theories by unifying gauge couplings at the grand unified
scale~\cite{uni}.
However, the new particles are not yet to be seen.
Neither the LEP nor the Tevatron is successful so far in their attempts to
discover these particles, and the search attempts have already exceeded the
$Z$ boson mass scale.
The SUSY extension which is invoked to explain the
electroweak scale seems to require most of the superpartners above
the electroweak scale.
In the SUSY breaking models mediated by minimal supergravity (mSUGRA)~\cite{sugra,sugra1},
where the squarks and sleptons masses are unified at the GUT scale, the average
stop mass scale is about 1 TeV or above.
Question now arises regarding the
justification of the heavy superpartner masses.
Can this little hierarchy between the $Z$ boson mass and the superpartner
masses be rationalized in these models?


We need to understand first the relation between
the SUSY breaking masses and the $Z$ boson mass.
The electroweak symmetry breaking relates the SUSY breaking mass scale to the $Z$ boson mass, $M_Z$.
At the tree level, the  minimization of the Higgs potential gives rise to
$M_Z^2/2\simeq -m_{H_u}^2-\mu^2$ in the large $\tan\beta$ limit, where $m_{H_u}$ is the
SUSY breaking mass for up-type Higgs boson, and $\mu$ is the Higgsino mass
which is the coefficient of the bilinear term
in the superpotential.
Since $|m_{H_u}|$ is of the order of the stop mass scale,
the natural expectation is that $M_Z$ is as large as stop mass, unless there is
a cancellation.
One can quantify the amount of cancellation by a sensitivity function
and finds that
smaller $\mu$ (and therefore small $|m_{H_u}|$) is needed \cite{nath}.
One can then conclude that the hierarchy between the SUSY breaking mass scale
 and $Z$ boson mass
is not preferred unless $\mu^2$ and $m_{H_u}^2$
are related in a given SUSY breaking model.
This  is called naturalness of the electroweak symmetry breaking.
%

In order to determine the
possible location of the SUSY breaking mass scale,
we need to go back to the origin of the SUSY models.
The SUSY models are expected to arise from well motivated string theory.
String theory has many vacua and
one expects to have wide range of possibilities of the SUSY model
parameters in these vacua \cite{Susskind:2003kw}.
The SUSY parameters can be the vacuum expectation values (VEVs) of
the moduli fields.
Many of these vacua can give rise to the SUSY
extension of the SM where the electroweak symmetry is broken.
One can then ask about the distribution of the model parameters in these
vacua once the requirement is made that the electroweak symmetry has
to be broken.
Can one understand the hierarchy between the SUSY breaking mass
scale and the $Z$ boson mass from the distribution of the model parameters?
One can also ask whether the same conclusion as naturalness
holds if a distribution function of the $|M_Z/m_{H_u}|$ hierarchy is considered.
The distribution functions are needed more than the sensitivity function
in the context of statistics of vacua.

So far we did not include the features of radiative symmetry
breaking~\cite{Inoue:1982pi} in our discussion.
Since the theory is not finite, one needs to care about large log correction
in the symmetry breaking conditions.
The SUSY breaking mass squared, $m_{H_u}^2$, is driven to be negative
at low energy by the renormalization group flow
and that leads us to satisfy the electroweak symmetry breaking condition.
The radiative symmetry breaking connects the stop mass scale to
the $Z$ boson mass in the following way.
The symmetry breaking condition
(i.e., $-m_{H_u}^2-\mu^2 >0$) is satisfied at a scale $Q_0$.
The tree-level $Z$ boson mass,
$M_Z^2 (Q)\simeq -2 (m_{H_u}^2+\mu^2)(Q)$, depends on the
renormalization scale, $Q$.
The proper $Z$ boson mass is given approximately
at the averaged stop mass scale $Q_{\tilde t}$
where the correction from 1-loop Higgs potential is negligible.
Those two scales, $Q_0$ and $Q_{\tilde t}$, are unrelated in general,
and
the electroweak symmetry is broken when $Q_0>Q_{\tilde t}$.
Now expanding the tree-level $Z$ boson mass, $M_Z(Q)$,
around the scale $Q_0$, one gets
$M_Z^2\propto \ln(Q_{0}/Q_{\tilde t})$.
Consequently,
the scales $Q_0$ and $Q_{\tilde t}$ need to be close by
when the average stop mass is about 1 TeV.
So instead of looking for a reason to explain the
smallness of $|M_Z/m_{H_u}|$, we need to understand
the proximity of the  scales $Q_0$ and $Q_{\tilde t}$.
We therefore direct our investigation to the distribution of the scales
$Q_0$ and $Q_{\tilde t}$ in the context of the statistics of vacua.
%

We consider the distribution of the hierarchy
between $M_Z$ and $Q_{\tilde t}$, and determine the distribution function
assuming that the any SUSY breaking vacuum is equally probable.
We determine whether the proximity of $Q_0$ and $Q_{\tilde t}$
is natural in a large number of vacua.
If this closeness is enough probable in the landscape of the
electroweak symmetry breaking vacua, the hierarchy between the
SUSY breaking scale and $M_Z$ can easily be rationalized
when $Q_0$ is at TeV scale due to a model parameter.
It is also interesting to determine the sensitivity
function of the $Q_0/Q_{\tilde t}$ hierarchy and compare it with
the distribution function.
We also determine  the probability of a given hierarchy between the
$Q_{\tilde t}$ and $M_Z$ in mSUGRA  model and show
90\% and 95\% probability contour in the experimentally allowed parameter space.
The average of the $\ln (Q_0/Q_{\tilde t})$ is considered
in a recent reference~\cite{Giudice:2006sn} in the
context of multiple vacua and it was shown that the $Q_0$ should be close to
$Q_{\tilde t}$ scale.
We propose to use the probability function
to describe the amount of little hierarchy in this paper.

It is not only interesting to investigate the distribution of
$Q_{\tilde t}$
to understand the little hierarchy,
but also important
to investigate the size of other parameters,
 especially the size of $\mu$, which is claimed to
be small for naturalness.
To investigate the size of $\mu$,
there is another important scale $Q_H$ in addition to the scales $Q_0$ and $Q_{\tilde t}$.
The $Q_H$ is the scale where $m_{H_u}^2$ becomes negative.
By definition, $Q_{\tilde t}< Q_0 < Q_H$ for the electroweak symmetry breaking vacua.
%
%
The hierarchy between $Q_0$ and $Q_H$
determines the preferred  size of $\mu$ and
therefore the size of $\mu$ can be understood for the distribution of $Q_0/Q_H$.
To obtain ``natural vacua" (or small $\mu$),
all three scales need to be close by.
It is also interesting to inquire about whether
there are lots of natural vacua
among the landscape of electroweak breaking vacua
varying the model parameters.


The paper is organized as follows. In section 2, we address the
little hierarchy problem.
In section 3, we discuss the conditions of the electroweak
symmetry breaking and determine the sensitivity function of little hierarchy.
In section 4, we describe the little hierarchy problem
in the landscape of electroweak vacua and  determine the probability
function of little hierarchy.
We also determine the 90\% and 95\% probability contours
in the mSUGRA model including the experimental constraints.
In section 5, we discuss the landscapes of different scale
associated with the electroweak symmetry breaking vacua and study the
possible size of $\mu$, and section 6 contains our conclusion.


\section{Little Hierarchy Problem}

The little hierarchy problem is often described by using
a sensitivity function \cite{nath}.
One can quantify the fine-tuning in the minimization condition
of Higgs potential by the sensitivity function
and concludes that small Higgsino mass $\mu$ is needed
for natural electroweak symmetry breaking.
However, distribution functions are more appropriate
rather than the sensitivity function
in the context of statistics of vacua.
In this section, we discuss the distribution
function of the hierarchy for the tree-level condition
to see whether the same conclusion also holds.


The tree-level Higgs potential is given as
\begin{equation}
V = m_1^2 | H_d |^2 + m_2^2 | H_u |^2 + (m_3^2 H_d \cdot H_u + h.c.)
+ \frac{g_2^2 + g^{\prime 2}}{8} (| H_d |^2 - | H_u |^2)^2
+ \frac{g_2^2}2 (H_u^\dagger H_d)(H_d^\dagger H_u)\,.
\label{Higgs-potential}
\end{equation}
The quartic term is obtained by $D$-term and thus
the coupling is related to the gauge couplings.
The quadratic terms are given by SUSY breaking Higgs masses,
$m_{H_d}^2$ and $m_{H_u}^2$, Higgsino mass $\mu$
and SUSY breaking bilinear Higgs mass $B\mu$ :
$m_1^2 = m_{H_d}^2 + \mu^2$, $m_2^2 = m_{H_u}^2 + \mu^2$
and $m_3^2 = B \mu$.

Minimizing the Higgs potential by Higgs VEVs 
 ($v_d = \langle H_d^0 \rangle$,
$v_u = \langle H_u^0 \rangle$),
we obtain
\begin{equation}
\frac{M_Z^2}2 =  \frac{m^2_{1}- m^2_2 \tan^2\beta}{\tan^2\beta -1}\,,
\qquad
\sin 2\beta = \frac{2 |m_3^2|}{m^2_{1} + m^2_{2}}\,,
\label{minimal}
\end{equation}
%
%
where $\tan\beta = v_u/v_d$.
The conditions of electroweak symmetry breaking at tree-level are
\begin{eqnarray}
&& m_1^2 \,m_2^2 < (m_3^2)^2 \,, \label{condition-1}\\
&& m_1^2 + m_2^2 > 2 |m_3^2| \label{condition-2} \,,
\end{eqnarray}
which corresponds to the conditions $M_Z^2 > 0$ and $\sin2\beta<1$ in Eq.(\ref{minimal}).
The second condition is obtained by the stabilization of the Higgs potential
along the flat direction, $|v_u| = |v_d|$.
The $Z$ boson mass can be expressed as
\begin{equation}
\frac{M_Z^2}2 = -\mu^2 + \frac{m^2_{H_d}- m^2_{H_u} \tan^2\beta}{\tan^2\beta -1}
\simeq -\mu^2 - m_{H_u}^2\,.
\label{Zboson}
\end{equation}

In the radiative electroweak symmetry breaking scenario \cite{Inoue:1982pi},
 the condition Eq.(\ref{condition-1}) is
satisfied at the weak scale by renormalization group equation (RGE).
The SUSY breaking scalar mass squared for up-type Higgs,
$m^2_{H_u}$, is driven to a negative value by large top Yukawa
coupling.
%
%
%
Naively, $-m_{H_u}^2$ is of the same order as the stop and gluino
masses at weak scale (especially, when the SUSY
breaking scalar masses are assumed to be universal), and
consequently, the $Z$ boson
mass is of the same order as the SUSY particles.
The colored particles are expected to be heavier than sleptons, wino
and bino due to the RGE effects using naive boundary conditions at
the Planck or the GUT scale.
In other word,
the uncolored SUSY particles should have been observed in LEP2 experiment.
%
%
Non-observation of the uncolored superparticles leads stop and gluino
masses to be much heavier than the $Z$ boson mass (especially when
the gaugino masses are unified at the GUT scale). Moreover, the lightest Higgs
mass bound ($m_h > 114.4$ GeV) pushes up the stop mass or the trilinear
scalar coupling for stop $A_t$.


Surely, there is a freedom of cancellation in
Eq.(\ref{Zboson}), and there is no problem with the electroweak
symmetry breaking even if SUSY particles are much heavier than the mass of the $Z$
boson.
%
However, the cancellation seems unnatural as can be seen in the  following
discussion.

The sensitivity function to measure the fine-tuning is defined as
\cite{nath}
\begin{equation}
\Delta[f(x)] \equiv \left|\frac{\partial \ln f}{\partial \ln x}\right|^{-1}.
\end{equation}
When $\Delta[f(x)]$ is small, the function $f$ is sensitive to $x$
and the degree of fine-tuning is large. The sensitivity for the $Z$
boson mass is calculated from Eq.(\ref{Zboson}) as
$\Delta[M_Z(\mu)]= M_Z^2/(2\mu^2)$. The $\mu$ parameter needs to be
small to generate less sensitivity. This is the usual naturalness
statement. When
\begin{equation}
M_H^2 \equiv (m_{H_d}^2 - m_{H_u}^2 \tan^2\beta)/(\tan^2\beta-1)
\end{equation}
is much larger than the $Z$ boson mass, the fine-tuning is severe.
For example, when $M_H= 500$ GeV, the sensitivity $\Delta[M_Z(\mu)]$
is about 2\%.

In order to describe the SUSY parameters in terms of the statistics of vacua, in
this paper, we suggest that the distribution function and the
probability function are more appropriate rather than the
sensitivity function.

Let us calculate the distribution function of
the $M_Z$-$M_H$ hierarchy ($r_H\equiv M_Z/M_H$).
%
Assume that any Higgsino mass $\mu$ is equally probable ($D[\mu]={\rm const}$).
Then one obtains the distribution function of $r_H$ as
\begin{equation}
D[r_H] = \frac{r_H}{2\sqrt{1-\frac{r_H^2}2}} \,,
\label{distribution-rH}
\end{equation}
by using the relation
$dP = D[r_H] dr_H = D[\mu] d\mu$.
The distribution function is normalized to make
 $\displaystyle\int_0^{\sqrt{2}} D[r_H] dr_H =1$.
Since the distribution function looks different by measure,
 we should use the probability function 
by integrating the distribution function
to avoid a bias for the choice of measure.
The probability for $r_H>r_0$ is given
as
\begin{equation}
P[r_H>r_0] = \int_{r_H}^{\sqrt2} D[r_H] dr_H = \sqrt{1-\frac{r_0^2}2}.
\end{equation}
So, the probability for $M_H < 2 M_Z$ is calculated to be 93\%.
The probability for $M_H> 200$ GeV (500 GeV) is only about 5\% (1\%).

Since the $\mu$ parameter is complex in general,
the proper distribution function of $\mu$ may be $D[|\mu|^2] = {\rm const}$
(or $D[\mu] \propto \mu$)
if any complex value is equally probable.
In this case, the distribution function is $D[r_H] = r_H$
and the probability function is $P[r_H < r_0] = r_0^2/2$.
The probability function can be written as $M_Z^2/(2M_H^2)$
which naively corresponds to $\Delta[M_Z(\mu)]$.
The probability for fine-tuning is relaxed than before:
The probability for $M_H> 200$ GeV (500 GeV) is about 10\% (2\%).
%
We note that
the complex $\mu$ does not mean CP violation
directly
since the phase of $\mu$ can be rotated out
by the field redefinition of Higgs fields
when $B$ parameter is real.

More generically,
the probability function is
\begin{equation}
P[r_H < r_0] = 1-\left(1-\frac{r_0^2}2\right)^{\frac{m}2},
\end{equation}
when $D[\mu^m] = {\rm const}$,
and
the probability can be written approximately $m {M_Z^2}/(4{M_H}^2)$
in the fine-tune region, which naively corresponds to
the sensitivity function of $\Delta[M_Z(\mu^{m/2})]$.
Therefore, naturalness statement holds
and the little hierarchy is not rationalized
even if we use the probability function
when $\mu$ is distributed.
%
%

%
%

There are mainly two directions to solve the little hierarchy problem.
In one direction one needs to select a suitable mass spectrum of SUSY particles at low energy.
For example, if squark, sleptons and wino are naturally heavier
than the SUSY breaking Higgs mass $M_H$ in a SUSY model,
the LEP2 experiments do not conflict with the fine-tuning.
In this case, a favorable SUSY breaking scenario will be
chosen such as mirage mediation model \cite{Choi:2005uz,Kitano:2005wc}.

The other direction is to reconsider the distribution of $\mu$ parameter
and to see what is a suitable parameter to distribute
in order to discuss the fine-tuning
in electroweak symmetry breaking.
In this paper, we investigate this direction.

\section{Conditions of Radiative Electroweak Symmetry Breaking}

In the previous section, we only study tree-level conditions,
Eq.(\ref{minimal}), which do not include the conditions that the
symmetry breaking is radiatively induced. Let us describe  the
conditions of the radiative electroweak symmetry breaking.

As we discussed, it seems that the fine-tuning is needed in Eq.(\ref{Zboson})
and the fine-tuning has less probability.
At what scale do we need the fine-tuning?
Since the mass parameters are running, we have to fix the scale where
we need fine-tuning.
The minimization conditions, Eq.(\ref{minimal}),
are given in the tree level for a given scale $Q$.
There exists a 1-loop corrected potential \cite{Coleman:1973jx},
\begin{equation}
\Delta V^{\rm 1-loop} = \frac1{64\pi^2} \sum_J
 (-1)^{2J} (2J+1) m_J^4 \left(\ln \frac{m_J^2}{Q^2} - \frac32\right)\,,
\end{equation}
where $J$ is a spin of the matter. Since $m_J$ depends on the Higgs
VEVs, we need to include the derivatives of 1-loop potential in
minimization of the Higgs potential. We can use a scheme that the
scale $Q$ is chosen to make the derivatives of 1-loop potential
$\partial \Delta V/\partial v_{u,d}$ to be small \cite{Casas:1998cf}. One can
find that the scale is a geometrical average of the stop mass,
$Q_{\tilde t} \equiv (m_{\tilde t_1} m_{\tilde t_2})^{1/2}$. As a
result, the tree-level relations Eq.(\ref{minimal}) are
approximately satisfied at the scale $Q_{\tilde t}$, and the
electroweak symmetry breaking conditions
Eqs.(\ref{condition-1},\ref{condition-2}) need to be satisfied at
$Q_{\tilde t}$. Defining the scale where the electroweak symmetry is
broken (Eq.(\ref{condition-1}) is satisfied) as $Q_0$, and the
scale where  the stability condition Eq.(\ref{condition-2}) is violated as $Q_{\rm
st}$, we can obtain the window of radiative electroweak symmetry
breaking as
\begin{equation}
Q_{\rm st} < Q_{\tilde t} < Q_0 \,.
\label{radiative-condition}
\end{equation}
%

To express the above statements explicitly,
we will make the $M_Z$ function as (in large $\tan\beta$
for simplicity\footnote{
For general $\tan \beta$, we obtain
\begin{equation}
M_{Z}^2\cos^2 2\beta
 \simeq
\left(
\frac{d m_{2}^2}{d \ln Q} \sin^2 \beta +
\frac{d m_{1}^2}{d \ln Q} \cos^2 \beta
- \frac{d m_3^2}{d \ln Q}\sin 2\beta \right)
\ln \left( \frac{Q_0}{Q_{\tilde t}}\right)^{\!\!2}.
\end{equation}
})
\begin{equation}
M_Z^2 (Q) = 2 (-\mu^2 - m_{H_u}^2)(Q) = -2 m_2^2(Q)\,.
\label{Z-boson-func}
\end{equation}
The true $Z$ boson mass is given as $M_Z^2 \simeq M_Z^2 (Q_{\tilde t})$.
By definition, $M_Z^2(Q_0) = 0$.
Therefore, expanding the function around $Q_0$,
we obtain
\begin{equation}
M_Z^2 \simeq \left. \ln \frac{Q_{\tilde t}}{Q_0}
\frac{d}{d \ln Q} M_Z^2 (Q) \right|_{Q=Q_{\tilde t}}
=\left. \ln \left(\frac{Q_0}{Q_{\tilde t}}\right)^{\!\!2}
  \frac{d m_2^2}{d \ln Q}  \right|_{Q= Q_{\tilde t}}.
\label{Z-boson-log}
\end{equation}
The 1-loop RGE of $m_2^2 = m^2_{H_u}+\mu^2$ is given in an appropriate notation as
\begin{eqnarray}
8\pi^2 \frac{d m_{H_u}^2}{d \ln Q}
&=& 3
 \left(y_t^2 (m^2_{\tilde t_L} + m^2_{\tilde t_R} + m_{H_u}^2) + A_t^2\right)
 -  (g^{\prime2} M_1^2 +3 g^2 M_2^2) + \frac12 g^{\prime 2} S \,,
\label{RGE}
\\
8\pi^2 \frac{d \mu^2}{d\ln Q}
&=& (3 y_t^2+3y_b^2+y_\tau^2 - g^{\prime2} -3 g_2^2) \mu^2 \,,
\end{eqnarray}
where $S$ is a trace of scalar masses with hypercharge weight.
Approximately, we obtain
\begin{equation}
M_Z^2  \simeq \frac{3}{8\pi^2}
\left( m_{\tilde t_L}^2 + m_{\tilde t_R}^2 + A_t^2 \right)
 \ln \left(\frac{Q_0}{Q_{\tilde t}}\right)^{\!\!2} ,
\label{Zboson2}
\end{equation}
neglecting gauge couplings $g^\prime$, $g_2$, and bottom, tau Yukawa couplings
$y_b$, $y_\tau$.

%
%
%

\begin{figure}[t]
 \center
 \includegraphics[viewport = 20 20 280 225,width=8.5cm]{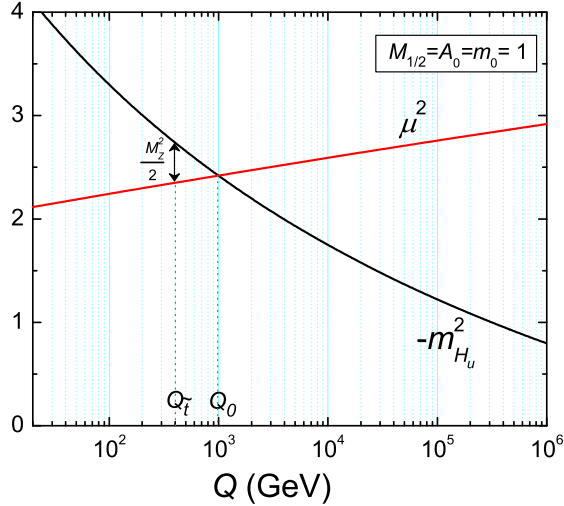}
 \caption{
RGE evolutions of $\mu^2$ and $-m_{H_u}^2$ are plotted as functions of the scale $Q$(GeV).
The Higgsino mass $\mu$ is chosen to make $Q_0 = 1$ TeV.
The proper $Z$ boson mass is evaluated at the scale $Q_{\tilde t}$.
}
\end{figure}

The interpretation of Eq.(\ref{Z-boson-log}) is illustrated in
Fig.1.
In Fig.1, the SUSY breaking mass parameters in
mSUGRA are made to be dimensionless unit since the RGE
evolution does not depend on overall scale factor.
As it is obvious from the figure, the little hierarchy is
characterized by the smallness of the triangle. The little hierarchy
problem can be rephrased in terms of the  question why the size of
the triangle is small. There are two ways to make the triangle
small. The vertical tuning corresponds to the tuning of $\mu$
parameter with fixed $M_H$ as in the previous section. The $\mu$
parameter tuning is equivalent to tuning $Q_0$ after fixing
$Q_{\tilde t}$. The horizontal adjustment of the triangle
corresponds to the tuning of $\ln (Q_0/Q_{\tilde t})$.

%
%

We stress that
the smallness of $\mu$ is not crucial
for the little hierarchy 
due to the fact that
$m_{H_u}^2$ and $\mu^2$ are canceled at $Q_0$.
%
In that sense, the usual fine-tune quantity $\Delta[M_Z(\mu)] =
M_Z^2/2\mu^2$ does not play a key role to describe fine-tuning in
radiative electroweak symmetry breaking when $Q_0$ is fixed at a TeV
scale.
We point out that
the horizontal adjusting quantity
$\ln (Q_0/Q_{\tilde t})$
is more important to discuss the little hierarchy
in radiative electroweak symmetry breaking.
Actually, one can calculate the sensitivity function of $\Delta[M_Z(Q_{\tilde t})]$ to be
\begin{equation}
\Delta[M_Z(Q_{\tilde t})] \simeq
\frac{\ln \left(\frac{Q_0}{Q_{\tilde t}}\right)^{\!2}}{\left|
1-\ln \left(\frac{Q_0}{Q_{\tilde t}}\right)^{\!2}   \right|}\,.
\label{sensitivity-q}
\end{equation}
 If $\ln (Q_0/Q_{\tilde t}) \sim 0$, then $\Delta[M_Z(Q_{\tilde
 t})]$ is small and $Z$ boson mass is sensitive to $Q_{\tilde t}$.


The question is now whether $\ln (Q_0/Q_{\tilde t})^2 \alt O(1)$ is
natural. 
Apparently, there is no
reason that $Q_{\tilde t}$ is related to $Q_0$\footnote{
We comment that $\ln (Q_0/Q_{\tilde t})^2 \simeq 1$ is satisfied in
the scenario of no-scale supergravity \cite{nano} in which the scale
$Q_{\tilde t}$ is determined dynamically.}.
For example, let us
assume that all the mass parameters (including $\mu$ and $B$) to be
proportional to a single mass scale $M_S$. Namely, the mass
parameters in the model are written as $m_{\tilde Q}^2 = \hat
m_{\tilde Q}^2 M_S^2$, $m_{\tilde g} = \hat m_{\tilde g} M_S$, $A_u
= \hat A_u M_S$, $\mu = \hat \mu M_S$ and so on. The dimensionless
coefficients $\hat m_{\tilde Q}^2$, $\hat m_{\tilde g}$,
 $\hat A_u$ etc
are determined when we fix a SUSY breaking scenario.
The $\mu$ parameter is also proportional to SUSY breaking scale
in Giudice-Masiero mechanism \cite{gm}
in which the $\mu$-term is forbidden in the superpotential by a symmetry
and the Higgsino mass originates from the Higgs bilinear term in K\"ahler potential.
In this case, $Q_0$ does not depend on $M_S$
since RGEs are homogeneous differential equations.
The averaged stop mass $Q_{\tilde t}$, on the other hand, of course depends on $M_S$.
Since the scale $Q_0$ is determined radiatively,
it is hierarchically smaller than the grand unified scale or the Planck scale $M_P$.
Namely, the scale $Q_{\tilde t}$ is a dimensionful parameter,
while
the scale $Q_0$ is expressed
as $Q_0 \sim e^{- 4\pi^2t} M_P$
 by a dimensionless $O(1)$ parameter $t$.
How can those two scales be related? Actually, in any SUSY breaking
scenario, the determination of the coefficients and the overall
scale are completely different issues. Why $Q_0$ and $Q_{\tilde t}$
are so close to make $\ln (Q_0/Q_{\tilde t})^2 \alt O(1)$?  This is
an essential question of radiative electroweak symmetry breaking and
the origin of SUSY breaking. In
addition, when $Q_0< Q_{\tilde t}$, the electroweak symmetry does
not break. It is just like living on the  edge of a cliff.

\section{Landscape of Electroweak Symmetry Breaking Vacua}

Anthropic principle teaches us that
we need not worry about the fact that
$Q_{\tilde t}$ appears within the window for electroweak
symmetry breaking, Eq.(\ref{radiative-condition}).
So, the question is whether
the little hierarchy is natural among the
electroweak symmetry breaking vacua.
To see that, we examine the
landscape of the electroweak symmetry breaking vacua.

As we have mentioned in the previous section, let us assume that all
mass parameters are proportional to single SUSY breaking mass scale
such that
 $m_{\tilde Q}^2 = \hat m_{\tilde Q}^2 M_S^2$,
$m_{\tilde g} = \hat m_{\tilde g} M_S$, $A_u = \hat A_u M_S$, $\mu =
\hat \mu M_S$ and so on. Then, the radiative electroweak symmetry
breaking scale $Q_0$ does not depend on $M_S$ when the dimensionless
coefficients are fixed. On the other hand, $Q_{\tilde t}$ is naively
proportional to $M_S$. The SUSY breaking mass scale $M_S$ is
specified by the $F$-term of a SUSY breaking spurion field $X$, as
$M_S \propto |F_X|/M_P$. If any complex value of $F_X$ is equally
probable, the distribution function of $M_S$ is $D[M_S] \propto
M_S$.
Therefore,
as one of the simplest example,
we calculate the distribution function
when the distribution of $Q_{\tilde t}$ is proportional to
$Q_{\tilde t}$ after fixing  $Q_0$.

Now, we calculate the distribution function of the $Z$ boson and the
stop mass hierarchy using Eq.(\ref{Zboson2}).
The hierarchy $R_{\tilde t} \equiv M_Z/\bar m_{\tilde t}$
is given as
\begin{equation}
R_{\tilde t}^2 = \alpha\ln  \frac{Q_0}{Q_{\tilde t}}\,,
\label{R_t}
\end{equation}
where $\bar m_{\tilde t}$ is an averaged stop mass, $\bar m_{\tilde
t}^2 =(m_{\tilde t_L}^2+m_{\tilde t_R}^2)/2 $ and $\alpha$ is the
coefficient
\begin{equation}
\alpha \simeq \frac3{4\pi^2} \left(2+\left(\frac{A_t}{\bar m_{\tilde t}}\right)^2\right).
\end{equation}
%
Using the relation, $D[R_{\tilde t}]=D[Q_{\tilde t}] dQ_{\tilde t}/d R_{\tilde t}$,
we obtain the distribution function of $R_{\tilde t}$
as
\begin{equation}
D[R_{\tilde t}] = \frac4{\alpha} R_{\tilde t} \,
\exp\left({- 2 \frac{R_{\tilde t}^2}{\alpha}}\right),
\end{equation}
where we normalize the distribution function
to make $\displaystyle
\int_0^\infty D[R_{\tilde t}] d R_{\tilde t} = 1$.
When $Q_{\tilde t} \ll Q_0$, the stability condition will break,
but we neglect the condition
to calculate the distribution function
since the distribution for $Q_{\tilde t} \ll Q_0$ ($R_{\tilde t} \gg 1$)
is exponentially suppressed.

\begin{figure}[t]
 \center
 \includegraphics[viewport = 20 20 280 225,width=8cm]{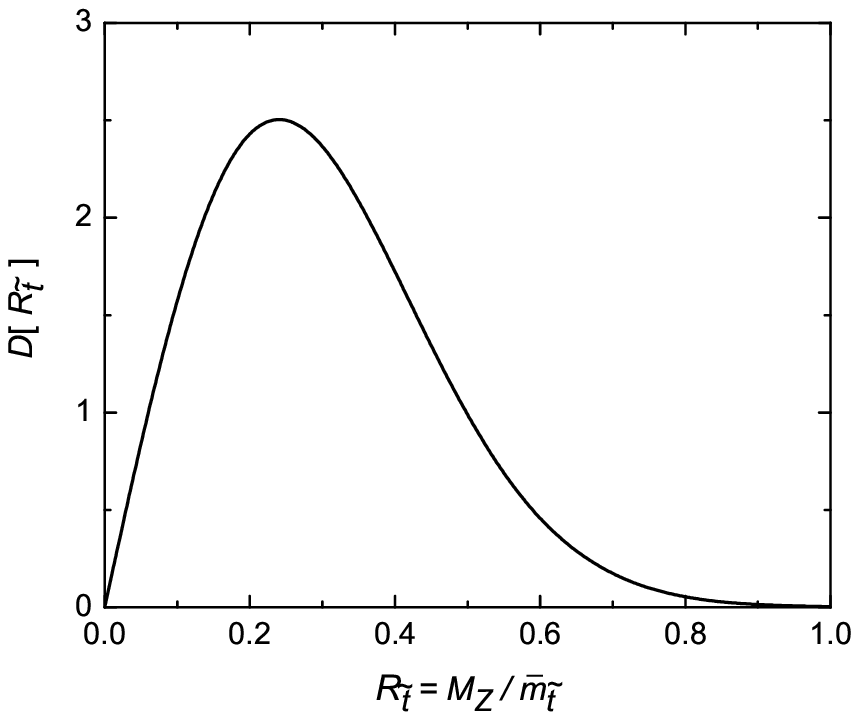}
 \includegraphics[viewport = 20 20 280 225,width=8cm]{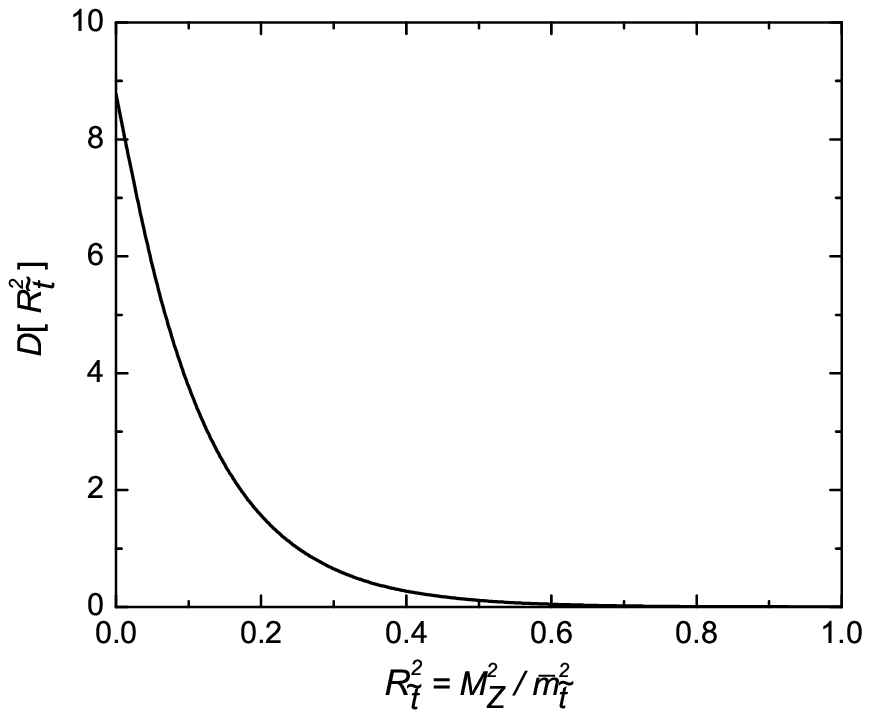}
 \caption{The distribution functions
$D[R_{\tilde t}]$ 
(left) and
$D[R^2_{\tilde t}]$ 
(right)
are shown.}
\end{figure}

We plot the distribution function $D[R_{\tilde t}]$ in the case
$A_t/{\bar m}_{\tilde t} = 1$ (namely $\alpha = 9/(4\pi^2)$) in
Fig.2. The peak of the distribution is $R_{\tilde t} =
\sqrt{\alpha}/2 \simeq 0.24.$ From the distribution, one can find
that a little hierarchy between the stop and the $Z$ boson masses
are probable among the electroweak symmetry breaking vacua. When we
look at  the distribution function of $R_{\tilde t}^2$ ($2R \cdot
D[R^2]= D[R]$),
\begin{equation}
D[R^2_{\tilde t}]
= \frac{2}{\alpha} \exp\left(-2 \frac{R_{\tilde t}^2}{\alpha}\right),
\end{equation}
it becomes more clear that there is a strong probability for $\ln
Q_0/Q_{\tilde t} \simeq 0$.

Usually, it is said that a small value of $\Delta[f(x)]$ is unwanted since $f$
is sensitive for $x$ and a fine-tuning is needed. In fact, for the
$\mu$ distribution 
in the section 2, the probability function naively corresponds to the sensitivity
function. However, we have encountered an example where the probability and
the sensitivity have different qualitative features. Namely, the
fine-tuning becomes most probable.
%


One may think that it looks awkward that fine-tuning is preferable.
However, it can happen when we consider a distribution.
Let us illustrate it for the distribution of $f(x)=a\ln (1/x)$.
Assume that any $x$ is equally probable for $0<x<1$.
Then $y=f(x)$ is distributed for $y>0$ and
the distribution function is obtained as
$D[y] \propto \exp(-y/a)$.
On the other hand, $\Delta[f(x)]= \ln (1/x)= y/a$.
Therefore, $y\sim 0$ is the most probable,
while the sensitivity function becomes zero at the point.
It can be understood intuitively from the semi-log graph
such as in the Fig.1.
The vertical lines are dense for larger values of horizontal logarithmic axis.
Surely, $y<0$ is more probable than $y>0$ if $x>1$ is allowed.
However, if we compare this example with our model of concern, $y<0$
corresponds to the vacua where the electroweak symmetry breaking would
not happen. Among the electroweak symmetry breaking vacua, therefore, the
fine-tuned vacua are more probable.
%
%
Furthermore, the distribution for less hierarchy is exponentially
suppressed due to the loop factor $\alpha$.

As seen in the Fig.2,
the shape of the distribution function looks different in different measures,
$dR_{\tilde t}$ or $dR_{\tilde t}^2$.
So it is better to use the probability function for a numerical quantity
of little hierarchy
instead of the distribution function.
The probability for $R_{\tilde t} > R_0$
is given by
\begin{equation}
P[R_{\tilde t} > R_0] = \int_{R_0}^\infty D[R_{\tilde t}] dR_{\tilde t}
= \exp \left(-2 \frac{R_0^2}{\alpha}\right).
\end{equation}
So, we obtain the probability for $\bar m_{\tilde t} > 2 M_Z \ (3 M_Z)$
is 89\% (62\%),
and
\begin{equation}
\bar m_{\tilde t} < (830 \ {\rm GeV})\ \sqrt{\frac{3}{2+(A_t/\bar m_{\tilde t})^2}} \,,
\label{prob-stop}
\end{equation}
at 90\% probability.

We have assumed that the distribution function of $M_S$
is proportional to $M_S$
because the SUSY breaking order parameter is complex.
When the order parameter is real in the case of $D$-term breaking,
the distribution function of $M_S$ is constant.
In general, if $n$ real components cooperate the overall SUSY breaking scale,
the distribution function is $D[M_S] \propto M_S^{n-1}$ (or
$D[M_S^n] = {\rm const}$),
and we obtain the probability function as
\begin{equation}
P[R_{\tilde t} > R_0]
= \exp \left(- n \frac{R_0^2}{\alpha}\right).
\label{prob}
\end{equation}



In Ref.\cite{Giudice:2006sn},
the authors use an average of $R_{\tilde t}^2$,
\begin{equation}
\langle R_{\tilde t}^2 \rangle
= \int_0^\infty R_{\tilde t}^2 D[R_{\tilde t}] dR_{\tilde t} = \frac{\alpha}{n}\,,
\end{equation}
to claim the closeness of the $Q_{\tilde t}$ and $Q_0$. The
probability that it is more hierarchical than the average is 63\%
($=1-1/e$), namely, it is about two times more probable rather than
that of less hierarchy. Therefore, we propose to use the probability
to describe the little hierarchy.

Since the sensitivity function is not
the proper quantity in the landscape picture distributing the overall
SUSY breaking scale,
we suggest to use the probability function Eq.(\ref{prob})
to characterize the little hierarchy.
We plot the 90\% and 95\% probability contours (in the case of $n=2$)
in the minimal supergravity
with $A_0= 0$ and $\tan\beta =10,40$.
To calculate the probability, we only distribute
the overall SUSY breaking scale
for each $m_0/m_{1/2}$ ratio.

As calculated in Eq.(\ref{prob-stop}),
the averaged stop mass is less than 1 TeV
at 90\% probability.
We can see from the figure that the
95\% probability region can be tested at the LHC and the future dark matter
detection experiments since the SUSY particle masses are not very large.
This region lies in the allowed parameter space. The parameter
space is constrained by the dark matter constraint~\cite{WMAP},
the lower limit on Higgs mass, LEP bounds
on SUSY particles~\cite{bound},
$b\rightarrow s\gamma$ bound~\cite{rare_decay} and
the muon $g-2$ data~\cite{muon}.

\begin{figure}[t]
 \center
 \includegraphics[viewport = 20 60 560 650,width=8cm]{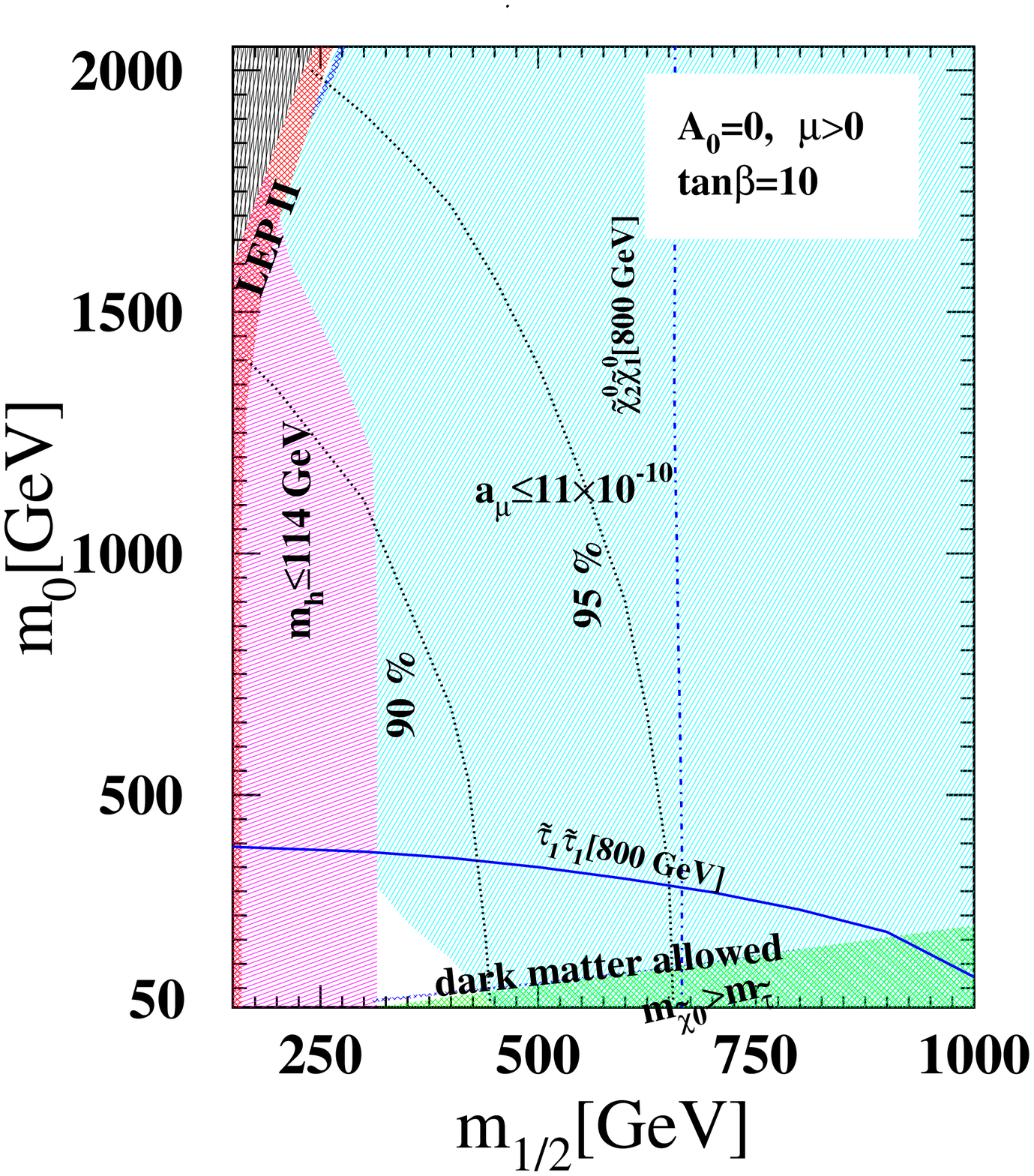}
 \includegraphics[viewport = 20 60 560 650,width=8cm]{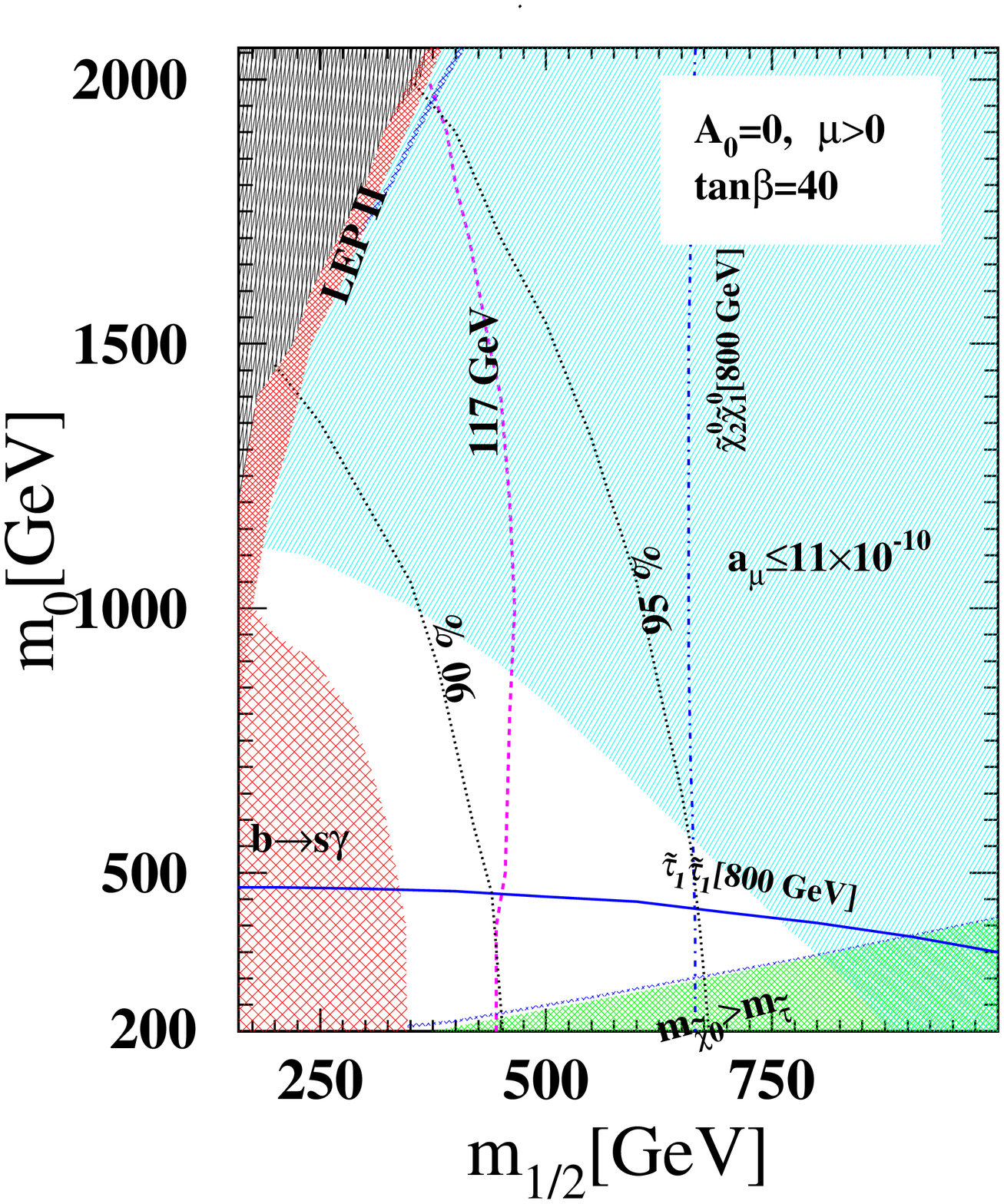}
 \caption{90\% and 95\%  probability contours (black lines) are  shown in the
 allowed parameter space for $\tan\beta=10$ (left) and
 $\tan\beta=40$ (right). The blue narrow bands are allowed by dark matter
 constraints. The lightest Higgs mass $m_H\leq 114$ GeV is in the pink shaded region.
 The red shaded region is disallowed by the LEP data.
 The lightest
 supersymmetry particle is charged in  the green region.
 $a_{\mu}\leq 11\times 10^{-10}$ in the light blue shaded region. The brick red hatched region
 obeys the $2.2\times 10^{-4}<{\rm Br}[b\rightarrow s\gamma]<4.5\times 10^{-4}$ constraint.
 The blue vertical and
 horizontal line show the ILC (800 GeV) reach in $\tilde\chi^0_2\tilde\chi^0_1$ and
 $\tilde\tau_1\tilde\tau_1$ final states. The black region is
 not allowed by radiative electroweak symmetry breaking. We use $m_t=172.7$ GeV for this graph.}
\end{figure}



\section{Several Landscapes in Minimal Supergravity}

In the minimal supergravity model \cite{sugra,sugra1}, the parameters are given
as ($m_0, m_{1/2}, A_0, \mu, B$). One usually uses $B$ to determine
$\tan\beta$, and uses $\mu$ to solve $M_Z$ using Eq.(\ref{Zboson}) at
the weak scale. So far, the parameter set is ($m_0, m_{1/2}, A_0,
\tan\beta, {\rm sgn}(\mu)$). In solving the equation for the $Z$
boson mass by $\mu$, one may need fine-tuning and the probability
of fine-tuning is small as we have seen in section 2.

In the landscape, as we have seen in the previous section,
the dimensionless parameters ($\hat m_0$, $\hat A_0$, $\hat \mu$, $\tan\beta$)
are given and one dimensionful scale $m_{1/2}$ is distributed,
where $\hat m_0 = m_0/m_{1/2}$, $\hat{A_0} = A_0/m_{1/2}$, $\hat \mu = \mu/m_{1/2}$.
%
The electroweak symmetry breaking scale $Q_0$
is the function of these four dimensionless parameters up to a cutoff scale,
$M_P$ or $M_{\rm GUT}$.
Once these four parameters are fixed,
$m_{1/2}$ is consumed to solve $Z$ boson mass
and a  fine-tuning may be needed.
However, the fine-tuning for the little hierarchy has enough probability
among the landscape of electroweak symmetry breaking vacua.

The difference in above two results depends on what parameter is distributed in the landscape.
In the first case, ($m_0$, $A_0$, $m_{1/2}$, $\tan\beta$) is fixed and $\mu$ is distributed,
and in the latter case, ($\hat m_0$, $\hat A_0$, $\hat \mu$, $\tan\beta$)
is fixed and $m_{1/2}$ is distributed.
In the landscape distributed by $\mu$,
the fine-tuning vacua is less probable,
and thus the small $\mu$ is demanded as in the usual naturalness statement.
We emphasize that the usual naturalness statement
is not necessarily applied in the landscape distributed by
$m_{1/2}$.

In the anthropic picture,
the landscape mostly prefers a little hierarchy
irrespective of $Q_0$.
We are interested in the vacua where
$Q_0$ is at TeV scale in our universe. %
In the parameter space where $Q_0$ is at TeV scale,
$\hat \mu$ is not necessarily small.
Actually,  $\hat \mu$
is almost determined irrespective of $\hat m_0$
 when $Q_0$ is at a TeV scale in the minimal supergravity
since $M_H^2$ has a focus point around the TeV scale \cite{focus}.
As a result, the $\mu$
parameter and the CP odd Higgs boson mass can be large with enough
probability in the landscape contrary to the usual naturalness
statement.

We have fixed the scale $Q_0$ in our discussion of   landscape
in the previous section.
What happens when $Q_0$ is also distributed?
To see that, let us study the distribution of $Q_0$
as well as the 
other parameters.

\subsection{Landscape of the scale $Q_0$}

Let us first see the landscape of the scale $Q_0$ distributed by $\mu$.
The scale is given as
\begin{equation}
\mu^2 (Q_0)= M_H^2 (Q_0) \simeq - m_{H_u}^2 (Q_0).
\end{equation}
The scale dependence of $\mu$ can be written as $\mu^2(Q_0) = \mu_0^2 I(Q_0)$.
Therefore, when any complex value of $\mu_0$ is
 equally probable ($D[|\mu_0|^2] = {\rm const}$),
the distribution of $t_0 \equiv \ln Q_0$
is
\begin{equation}
D[t_0] = D[\mu^2_0] \frac{d\mu_0^2}{dt_0} = c\, \frac{d}{dt_0} \frac{{M}_H^2(t_0)}{I(t_0)},
\end{equation}
where $c$ is a constant.
%
The $\mu$ parameter may be a function of moduli, e.g. $\mu = f(z)$.
Then the $\mu$ distribution is $D[\mu] = D[z] df^{-1}/d\mu$.
In the case where $D[\mu^m_0] = {\rm const}$,
we obtain
\begin{equation}
D[t_0] = c
\left(\frac{M_H^2}{I}\right)^{\frac{m-2}2} \frac{d}{dt_0} \frac{M_H^2}{I}.
\end{equation}

The maximal value of $Q_0$ is the scale where $M_H^2 =0$. We define this
 scale $Q_H$, namely $M_H^2(Q_H)=0$.
Surely, $Q_H$ does not depend on $\mu$ and the
overall mass scale. Since the RGE of $m^2_{H_u}$ becomes larger at
lower scale, the large $Q_0$-$Q_H$ hierarchy ($Q_0 \ll Q_H$) is more
probable for $m\geq2$.
 However, the little hierarchy between $Z$ boson mass and
SUSY breaking masses is not very probable as we have seen in section 2.

\subsection{Landscape of the scale $Q_H$}

How about the $Q_H$ landscape?
The scale $Q_H$ is function of $\hat m_0$, $\hat A_0$ (in the unit of $m_{1/2}$)
and $\tan\beta$.
For simplicity, let us choose $A_0 = 0$ and neglect $\tan\beta$ dependence.
We parameterize $\hat m_0 = \tan\theta_T$
and any $\theta_T$ ($0<\theta_T<\pi/2$) is equally probable.
The $\theta_T$ can be identified to the mixing between dilaton and moduli
which breaks SUSY.
The scale dependence of $m_{H_u}^2$
can be written as
\begin{equation}
m_{H_u}^2 (Q) = m_0^2 K_0(Q) + m_{1/2}^2 K_{1/2}(Q).
\end{equation}
Then the distribution function of the scale $t_H \equiv \ln Q_H$
where $m_{H_u} (Q_H)=0$
is
\begin{equation}
D[t_H] = c \, \sqrt{\frac{K_0}{K_{1/2}}} \frac{K_0}{K_0+K_{1/2}} \frac{d}{dt} \frac{K_{1/2}}{K_0}.
\end{equation}
In the minimal supergravity,
$Q_H$ has a maximal value $Q_H^{\rm max} \sim 10^{10}$ GeV for $\hat m_0 = 0$
and $K_{1/2}(Q_H^{\rm max}) = 0$,
and a minimal value $Q_H^{\rm min} \sim 5$ TeV (when $\tan\beta =40$)
for large $\hat m_0$
and $K_0(Q_H^{\rm min})=0$.
We plot the distribution function in Fig.4.
The scale $t_H$ does not have strong preference
except for the scales to be around the minimal and maximal values.
The probability around the minimal and maximal values arising
from the integration of the distribution function is not very large.
Therefore, in the landscape, we do not obtain a typical preference of the
hierarchy.

\begin{figure}[t]
 \center
 \includegraphics[viewport = 20 20 280 225,width=8.5cm]{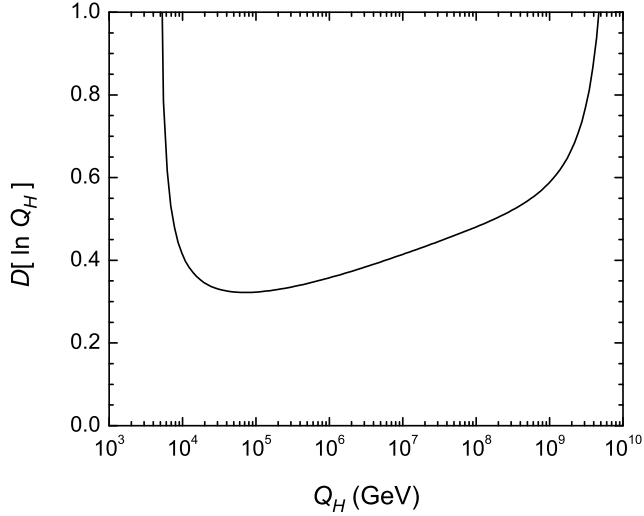}
 \caption{The distribution of $Q_H$ is shown when $\hat m_0 = m_0/m_{1/2}$ is distributed.
 $Q_H$ is the scale where SUSY breaking Higgs mass squared becomes negative.}
\end{figure}

\subsection{Landscapes of $Q_{\tilde t}$ and $Q_0$}

We distribute both $m_{1/2}$ and $\hat \mu$
with $D[m_{1/2}^2]=D[\hat \mu^2]={\rm const}$.
The distribution function is given
as $dP = D[m_{1/2}^2] D[\hat \mu^2] dm_{1/2}^2 d\hat\mu^2$.
We note that $d\mu$ and $d\hat\mu$ is the same measure up to normalization
when $m_{1/2}$ is fixed,
but if $m_{1/2}$ is also distributed, those two measures
 need to be distinguished.

When $m_{1/2}$ and $\hat \mu$ are distributed,
both $Q_{\tilde t}$ and $Q_0$ are distributed.
The distribution function of $t_0$ and $\tilde t \equiv \ln Q_{\tilde t}$
can be written as
$D[t_0,\tilde t] = c \, \theta(t_0-\tilde t)
\, e^{2 \tilde t} \frac{d}{dt}\frac{M_H^2}{I}|_{t=t_0}$,
where $\theta$ is a step function.
Defining $t_Z \equiv t_0 -\tilde t$ ($t_Z > 0$),
we obtain
\begin{equation}
D[t_0, t_Z] = c\, e^{-2t_Z} e^{2t_0} \left. \frac{d}{dt}\frac{M_H^2}{I}\right|_{t=t_0},
\end{equation}
and the distribution function can be decomposed
as $D[t_Z] \propto e^{-2t_Z}$
and $D[t_0] \propto e^{2t_0}
|\frac{d}{dt}M_H^2/I|$.
The hierarchy between the $Z$ boson and stop mass
is given as $M_Z^2/\bar m_{\tilde t} = \alpha t_Z$ as in Eq.(\ref{R_t}).
The
little hierarchy of $M_Z$-$\bar m_{\tilde t}$ is probable
from the distribution function $D[t_Z] \propto e^{-2t_Z}$
in the same way when we just use  $m_{1/2}$ distribution.

Since the large $t_0$ is strongly probable
due to the exponential factor in the distribution function
$D[t_0] \propto e^{2t_0}
|\frac{d}{dt}M_H^2/I|$,
the scale $Q_0$ is most probably
just below the maximal value of $Q_0$ which is the scale $Q_H$.
Therefore, all three scales, $Q_{\tilde t}$, $Q_0$ and $Q_H$ are close by.
Since $Q_0$ is just below the scale $Q_H$,
the $\mu$ parameter must be small by definition:
\begin{equation}
\mu^2(Q_0) \simeq \ln \frac{Q_0}{Q_H} \frac{d}{dt} M_H^2.
\end{equation}
In fact, the small $\mu$ is the most probable as one can see from
the distribution function which is calculated as
\begin{equation}
D[\mu^2(Q_0)] \simeq c
\left( 1+ \frac{\dot \mu^2}{|\dot M_H^2|} \right)
\exp \left(-2 \frac{\mu^2}{|\dot M_H^2|}\right),
\end{equation}
where dot represents for $t=\ln Q$ derivative.
Therefore,
this landscape mostly prefers the little hierarchy with
small Higgsino mass,
which is a demand from naturalness.
%
%
For our universe,
we have to choose the SUSY breaking scenario to make the $Q_H$ to be
TeV scale.
In the minimal supergravity,
$\hat m_0$ needs to be large to make $Q_H$ to be at the TeV scale,
which corresponds to the focus point solution \cite{focus}.
We stress that  naturalness is not required in this landscape,
but the naturalness vacua are most probable.

We remark that $Q_H$ must not be distributed
in this landscape,
otherwise SUSY breaking scale becomes just below the maximal value of
$Q_H$,
which is $10^{10}$ GeV in minimal supergravity.

\subsection{Summary of different Landscapes}

We have studied several landscapes to distribute parameters
($m_{1/2}$, $\hat \mu$, $\hat m_0$) in minimal supergravity.
The important scales to describe the landscape of
electroweak symmetry breaking vacua
are
stop mass $Q_{\tilde t}$,
symmetry breaking scale $Q_0$ (at $\mu^2 = M_H^2$),
and the scale $Q_H$ where $M_H = 0$.
The scales $Q_0$ and $Q_H$ do not depend on the overall mass scale
which we choose $m_{1/2}$,
and $Q_H$ does not depend on $\hat \mu$.

We consider the following typical landscapes:
\begin{enumerate}

\item
If we distribute the overall mass scale that is chosen to be
$m_{1/2}$, we can obtain a little hierarchy with enough probability
irrespective of the sensitivity function. Naturalness (smallness of
$\mu$) is not necessary in the landscape.

\item
If we distribute $\mu$ and fix the overall scale, the little
hierarchy between $Z$ boson mass and SUSY breaking scale is not
probable, but the hierarchy between $Q_0$-$Q_H$ is probable i.e., a
large $\mu$ could exist.

\item
If we distribute only $\hat m_0$, we do not obtain
any particular probable hierarchy.

\item
If we distribute both $m_{1/2}$ and $\hat \mu$, it is probable that
all three scales are close. Therefore, naturalness vacua with little
hierarchy is the most probable. The SUSY breaking scenario needs to
be fixed to make $Q_H$ (or $Q_H^{\rm max}$ if it is distributed) to be TeV scale
in our universe.
In the mSUGRA model, $\hat m_0 \simeq O(10)$ is required.
One can consider specific SUSY breaking models
such as
in Ref.\cite{Nomura:2006sw,Dermisek:2006ey}.

\end{enumerate}

\section{Conclusion}

The absence of the SUSY signals at LEP and Tevatron has pushed up
the SUSY particle mass scale compared to
the $M_Z$ scale.
The colored SUSY particles are now around 1 TeV scale in the mSUGRA models and therefore
created a little hierarchy between this scale and the $Z$ boson mass scale.
It is said that naturalness of the electroweak symmetry breaking requires
the smallness of the Higgsino mass $\mu$.

We investigated this situation
in the context of landscape of electroweak symmetry breaking vacua.
We include radiative symmetry breaking and found that in
order
to obtain a little hierarchy between $Z$ boson mass and SUSY breaking scale
with enough probability,
we need to distribute the overall SUSY breaking mass scale.
%
%
In this landscape, the naturalness (small value of $\mu$) is not required.
The Higgsino mass $\mu$ can be large or small
and
the scale $Q_0$ where the electroweak symmetry breaking conditions are satisfied
 needs to be chosen around 1 TeV 
 in our universe as one of the vacua in the landscape of little hierarchy.
In this scenario, the SUSY breaking mass is preferred to be
just below the scale $Q_0$,
%
in the electroweak symmetry breaking vacua and therefore the
  little hierarchy can be rationalized.

If $\mu$ is also distributed along with the overall SUSY breaking mass, 
natural vacua (small $\mu$) is found to be probable
among the electroweak symmetry breaking vacua.
In this landscape,
the scale $Q_H$, where the SUSY breaking Higgs mass squared turns negative,
has to be selected at a TeV scale in our universe
by choosing a SUSY breaking scenario.

If we only distribute $\mu$, the little hierarchy is less probable,
and the naturalness is demanded
as usually discussed.

We note that the landscape with overall scale distribution
supports the little hierarchy with enough probability,
but do not support huge hierarchy between SUSY breaking scale
and the $Z$ boson mass, such as split SUSY \cite{Arkani-Hamed:2004fb}
or non-SUSY standard model at low energy where all SUSY particles are decoupled.
Actually,
the stop mass is less than 3 TeV at 99\% probability.
We also comment
that  the vacua
with all scalar particles (including Higgs fields) and gauginos being decoupled
are enormously probable rather than low energy SUSY vacua
in this landscape picture.
The proper statement is that
the little hierarchy is mostly probable
among the low energy SUSY vacua with radiative electroweak symmetry breaking
by Higgs mechanism.


\end{document}